\documentclass[aps,twocolumn, showpacs]{revtex4}
\usepackage{graphicx}
\usepackage{dcolumn}
\usepackage[french]{babel}
\usepackage[latin1]{inputenc}
\usepackage[T1]{fontenc}

\begin{document}


\title{Magnetic patterning of (Ga,Mn)As by hydrogen passivation }

\author{L. Thevenard} \email{laura.thevenard@lpn.cnrs.fr}
\author{A. Miard}
\author{L. Vila}
\author{G. Faini}
\author{A. Lema\^{\i}tre}

\affiliation{Laboratoire de Photonique et de
Nanostructures CNRS, Route de Nozay, 91460 Marcoussis, France}

\author{N. Vernier}
\author{J. Ferr\'{e}}
\affiliation{Laboratoire de Physique des Solides CNRS (UMR 8502), Universit\'{e} Paris-Sud, 91405 Orsay, France}

\author{S. Fusil}
\affiliation{Unit\'{e} Mixte de Physique CNRS-Thal\`{e}s (UMR 137)
RD 128, 91767 Palaiseau Cedex,  France }


\date{\today}

\begin{abstract}

We present an original method to  magnetically pattern thin layers of (Ga,Mn)As. It relies on local hydrogen passivation to significantly lower the hole density, and thereby locally suppress the carrier-mediated ferromagnetic phase. The sample surface is thus maintained continuous, and the minimal structure size is of about 200 nm. In micron-sized  ferromagnetic dots fabricated by hydrogen passivation on perpendicularly magnetized layers, the switching fields can be maintained closer to the continuous film coercivity, compared to dots made by usual dry etch techniques. 

\end{abstract}


\maketitle

	Active research on the diluted magnetic semiconductor (DMS) Ga$_{1-x}$Mn$_{x}$As is now quickly catching up with the extensive knowledge accumulated on metallic ferromagnets. However, envisioning future applications for this material  presents several challenges. In addition to raising the Curie temperature \cite{wang04b}, it is necessary to understand the mechanisms of the magnetization reversal in single-domain systems. Novel magnetic behaviors are indeed expected to arise, where demagnetizing, edge, and finite volume effects come into play, as the dimensions become comparable to the typical  magnetic lengths (domain width, domain wall (DW) width, and exchange length).

	Fabrication of DMS microstructures is usually done using dry etch techniques \cite{Ohno00,tang06, yamanouchi:096601,tang04}. While having provided experimental results on DW velocity \cite{yamanouchi:096601,tang06}, this approach may however create an edge roughness along thin patterns. These irregularities are likely to grip the DW, inducing high depinning fields and process-dependent measurements.
	
	In this paper, we present a novel magnetic patterning technique, designed to circumvent this inconvenience. Taking advantage of the carrier-induced ferromagnetism in Ga$_{1-x}$Mn$_{x}$As, we use atomic hydrogen to form electrically inactive complexes with the magnetic impurities, thereby suppressing the ferromagnetic phase \cite{Lemaitre05,Goennenwein04,Bouanani03}.  By depositing a mask before the passivation, we pattern a perpendicularly magnetized layer into micron-sized ferromagnetic dots. After the hydrogenation process, the mask is removed. Only the zones shielded from the plasma remain ferromagnetic, while the rest of the layer is paramagnetic. Interestingly, this patterning route is reversible, since a low temperature anneal can restore the properties of the initial, non-patterned layer, by breaking the (Mn,H) complexes \cite{Thevenard05,thevenard07}. The main advantages of this purely diffusive process are to pattern reversibly a (Ga,Mn)As layer, contrary to ion implantation or etching, and to maintain the continuity of the film, which allows near field microscopy investigations and smoothing of border effects. Note that this patterning process had been proposed in earlier communications \cite{MMM07,dubon07,Brandt03}.

	The sample was grown by Molecular Beam Epitaxy following the procedure described in detail in Ref. \cite{Ferre05}. It consists of a 50~nm  Ga$_{0.93}$Mn$_{0.07}$As layer grown on a relaxed Ga$_{0.90}$In$_{0.10}$As buffer, with a Curie temperature  of 118~K after annealing and an effective perpendicular anisotropy field of $H_{u}\approx 3500~Oe$ at 2 K. In a previous paper \cite{Ferre05}, we showed on a very similar film that the magnetic easy axis was along [001] at all temperatures. Rare, growth-induced defects prevented the periodic striped arrangement of the domain structure expected for a perfect uniaxial ferromagnetic film. However, domains were homogeneous and 20 $\mu$m wide at 80~K, making them suitable for patterning single-domain magnetic elements.

	The patterning was done in the following way: after careful desoxydation, a titanium mask was first deposited by e-beam lithography lift-off. A 40 nm thickness was sufficient to prevent the hydrogen atoms from entering the layer. The mask consisted of three 200 $\times$ 200 $\mu$m arrays of dots, with sizes (and spacings): 10~$\mu$m (10~$\mu$m), 5~$\mu$m (10~$\mu$m) and 1~$\mu$m (5~$\mu$m). The sample was then exposed to a hydrogen plasma during 2~h; it will be referred to in the rest of the paper as "sample $H$". Transport measurements showed an increase of the sheet resistivity by two orders of magnitude, confirming that the layer had indeed been passivated. For comparison, another set of arrays, "sample $E$", was then processed by etching, using the same layer and mask. Ion Beam Etching (IBE) at a 20$^{\circ}$ incidence angle was used to ensure 80 nm high, vertical sidewalls. Finally, the metallic mask was removed off both samples by a diluted HF solution. Here, we emphasize that the surface of  sample $H$ was then completely continuous, whereas sample $E$ was left structured. 

\begin{figure}[!h]
 	\includegraphics{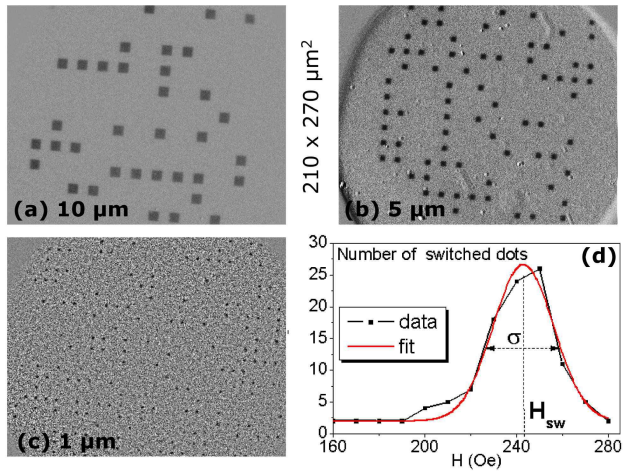}
	\caption{\small{ (a)-(c) : Kerr microscopy snapshots (T=85 K) during the magnetization reversal of  dots patterned by  hydrogen passivation. A 5 s field pulse of H=100, 100, and 120 Oe was applied to dot arrays of 10, 5 and  1 $\mu m$. The contrast is low for the latter considering the resolution of our PMOKE microscope ($\lambda=1~µm$). (d) Gaussian fit of the number of reversed 10 $\mu m$ dots, to extract a mean switching field $H_{sw}=243\pm 6~Oe$ and a field dispersion $\sigma=26\pm 1~Oe$ (T=20 K).}}
	\label{fig:sampleH}
\end{figure}

The magnetization reversal was then investigated between 4 K and 100 K by polar magneto-optical Kerr (PMOKE) microscopy, using a set-up described in Refs.  \cite{Ferre2001, Hubert1998}. This technique  yields an excellent contrast for Ga$_{1-x}$Mn$_{x}$As thin films \cite{Ferre05} and is particularly well suited to the study of our samples since the signal is sensitive to the perpendicular magnetization component. Snapshots were taken in zero field after applying $5~s$ pulses of positive magnetic field, and substracted from a reference image taken at negative saturation, to highlight the magnetic contrast between up- and down- magnetized areas. 

Field pulses of increasing amplitude were applied until complete magnetization reversal of the array. Images taken on sample $H$ during the reversal (Fig. \ref{fig:sampleH}a-c) clearly show that the passivation technique was successful in patterning  magnetically the layer into 1, 5 and 10 $\mu m$ structures. The arrays of etched dots follow a  comparable magnetic behavior (not shown here). As expected from the low saturation magnetization, we moreover verified that dipolar or exchange interactions between dots played a minimal role, since the number of reversed dots followed exactly an independent-dot probability law \cite{jamet98}.

\begin{figure}[!h]
		\includegraphics[width=7.5cm]{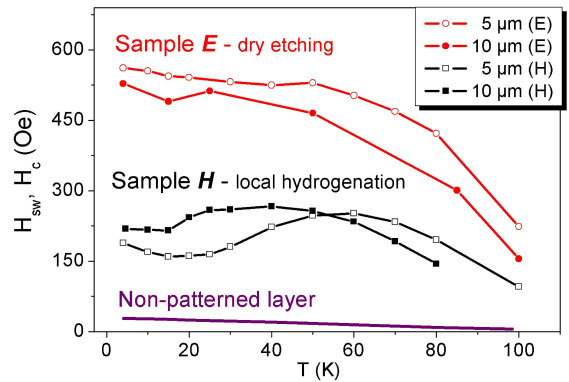}
	\caption{\small{ Mean switching fields of 10 $\mu m$ (full symbols) and 5 $\mu m$ (empty symbols) dots patterned by hydrogenation (black  squares) and by etching (red circles). The coercive field $H_{c}$ of the non-patterned layer was also measured by Kerr effect. Above 100 K, the contrast was too weak to provide reliable measurements.}}
	\label{fig:IBE_hyd}
\end{figure}

By plotting the number of dots that switched between two consecutive field pulses and fitting it to a gaussian, we then extracted a mean switching field, $H_{sw}$, and the corresponding field dispersion $\sigma$ for a given dot array and temperature (Fig. \ref{fig:sampleH}d). $H_{sw}$ values obtained by this procedure were reproducible within $\pm$ 20~ Oe. The resulting switching fields for 5 and 10 $\mu m$ dot arrays on samples $H$ and $E$ are shown in Fig. \ref{fig:IBE_hyd}, along with the coercive field, $H_{c}$, of the non-patterned layer determined by PMOKE.

The mean switching fields of sample $E$ are systematically much higher than $H_{c}$, and decrease by about $10\%$ when the perimeter of the dot doubles. $H_{sw}$ varies monotonously from 550 to 200 Oe when the temperature increases from 4 K to 100 K, with a fairly large relative dispersion $\frac{\sigma}{H_{sw}}$ from 22$\%$ at 4~K to 30$\%$ at 100~K. Dot arrays of sample $H$ have a quite different behavior. Switching fields are at least twice smaller than those of sample $E$, while remaining above $H_{c}$. They are also better defined at all temperatures, with a relative dispersion $\frac{\sigma}{H_{sw}}$  from 10 (T=4~K) to 18$\%$ (T=100~K). Most remarkable is the non-monotonous evolution of $H_{sw}$ with temperature, reaching a maximum at $T_{max}$=40 K (resp. 60 K) for 10$\%$ $\mu m$ (resp. 5 $\mu m$) dots. This unusual behavior was reproducible after temperature cycling.

In both samples, no partly-switched dot was observed in the accessible field and time range. The switching fields are however five to ten times smaller than those expected for a single-domain, coherent reversal ($H_{sw}=H_{u}$), suggesting a nucleation/propagation reversal mechanism. On the non-patterned layer, the magnetization reversal was shown to be triggered at rare nucleation centers (1 $mm^{-2}$), before rapidly developing by  easy domain-wall propagation \cite{Ferre05}. Let us first consider the etched dots, sample $E$. Upon reducing the size of the system, the nucleation becomes more difficult since the probability of finding nucleating centers with low energy barriers is decreased by a simple geometrical effect \cite{jamet98}.  From the weak sensitivity of $H_{sw}$ on the area of etched dots, it can be inferred that another mechanism is certainly involved, such as nucleations initiated on the rough edge of the dot, as has already been observed on metallic microstructures \cite{deak00}. The perimeter reduction of the dot entails a decreased probabibility of finding a nucleating defect along the edge, and therefore leads to a higher $H_{sw}$. The thermal activation of the nucleation process is responsible for the decrease of switching fields with temperature. After nucleation, the reversal proceeds by fast propagation within the dot.

Another mechanism has to be invoked for sample $H$, similar to that evidenced in Pt/Co/Pt dots patterned by focused Ga ion beam irradiation \cite{aign98}. The hydrogen passivation over the mask results in a hole density gradient within a ring around the ferromagnetic dot, a gradual interface between the dot and the  paramagnetic (Ga,Mn)As:H matrix. This soft  magnetic zone is likely to be an easy nucleation region from which the magnetization reversal is initiated. The dispersion is also expected to be much thinner than in sample $E$, since the reversal by wall propagation inside the dot no longer depends on a statistical distribution of defects around the dot edge, but on a fairly smooth, soft magnetic interface, hence resulting in the lower values of  $H_{sw}$ and $\frac{\sigma}{H_{sw}}$ for sample $H$.

In the soft ring around the dot, the hole density is high enough to yield a ferromagnetic phase, but sufficiently low to induce an in-plane magnetization, as has been predicted by mean-field theories \cite{Dietl01a}, and shown experimentally \cite{titova05,Thevenard05}. This may be compatible with the unusual temperature dependence of $H_{sw}$, since the in-plane magnetization at low temperature and low hole density is expected to  flip out of the plane, along [001], with increasing temperature \cite{titova05,thevenard07}. The switching fields will then \textit{increase} since spins will be blocked up to the coercive field, instead of rotating continuously with the applied field, along a hard-axis-like curve. The behavior of $H_{sw}$ may therefore reflect the temperature evolution of the magnetic anisotropy in the ring around the ferromagnetic dot. 

\begin{figure}[!h]
		\includegraphics[width=8.5cm]{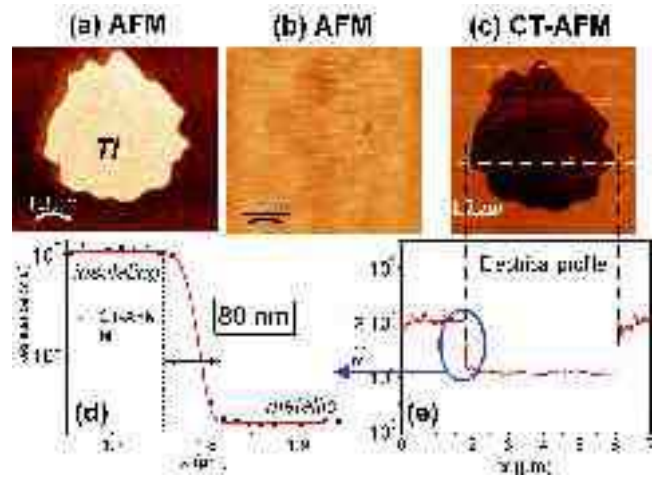}
	\caption{\small{ (a) AFM image of the mask used for the patterning. Note that its irregular shape is a non-intended lithography artefact. (b) AFM image after mask removal: the surface is continuous. (c) Conductive-Tip AFM scan after mask removal, revealing an electrical pattern exactly identical to the mask's. (d,e) The logarithmic profile evidences a two orders of magnitude decrease of the resistivity over about 80 nm. In this case, the characteristic diffusion length was estimated by an error function fit (line) to about 20 nm.}}
	\label{fig:CTAPL07}
\end{figure}

In order to estimate the width of the hydrogen diffusion front that constitutes a soft magnetic ring around the  dots, we used conductive-tip Atomic Force Microscope (CT-AFM) measurements to establish a resistivity mapping of the dot. In first approximation, we assumed it  to be proportional to the hydrogen density. This room-temperature technique associates a standard AFM with a voltmeter connected between the $p$-doped diamond tip (sample surface) and a back contact. Thus, it is preferable to work with conductive substrates. Ferromagnetic 4 $\mu m$ wide dots were therefore fabricated following the same technological process, on a 50 nm Ga$_{0.93}$Mn$_{0.07}$As layer grown over  GaAs:Be (Fig. \ref{fig:CTAPL07}a). 

 After removing the metal mask, a first regular AFM scan of the dot showed a smooth surface with a 2 nm roughness (Fig. \ref{fig:CTAPL07}b). The CT-AFM measurement then evidenced an electrical pattern on the layer identical to the shape of the mask (Fig. \ref{fig:CTAPL07}c). The profile taken along this scan confirmed that the resistivity increased by two orders of magnitude outside the dot (Fig. \ref{fig:CTAPL07}e). The interface profiles were then adjusted by a standard diffusion error function. This is a fairly crude approximation since it entails a 1-D diffusion process \cite{tuck88}. However, averaged over a dozen profiles, it  yielded a characteristic diffusion length of 27 nm, and a mean diffusion front width of 87 nm (Fig. \ref{fig:CTAPL07}d),  a value compatible with Kerr microscopy experiments done on small structures. This distance being larger than the DW width ($\approx$ 20 nm \cite{oszwaldowski06}) and the exchange length ($\approx$ 2 nm), a distinct ferromagnetic phase is established in a ring around the dot.
 
 In conclusion, we have patterned perpendicularly magnetized (Ga,Mn)As layers using a mono-atomic hydrogen plasma. Novel magnetization reversal phenomena arise from the presence of a soft magnetic interface between the ferromagnetic and paramagnetic regions, yielding smaller, and better defined  array switching fields than those in the structures made by dry etch. This technique seems ideal for performing DW velocity measurements in thin (Ga,Mn)As stripes, without edge roughness artefacts. At this stage of the study, the minimum structure size accessible by  this method was estimated around 160-200 nm, but it could reasonably be lowered by working with thinner layers that require shorter hydrogenation times. Moreover, it could be extended to other DMS, provided that their magnetic phase is carrier-mediated, and that undergoing hydrogenation indeed leads to their efficient passivation.

\begin{acknowledgments}
This work has been supported by the R\'{e}gion \^{I}le de France, the Conseil G\'{e}n\'{e}ral de
l'Essonne and through the Actions Concert\'{e}es Nanostructures DECORESS and
Incitative BOITQUAN. 
\end{acknowledgments}

\end{document}